\documentclass[pdftex,twocolumn,nopacs,arxiv]{svarxiv}  

\RequirePackage[T1]{fontenc}

\smartqed 
\pdfoutput=1
\RequirePackage{graphicx}
\usepackage{mathtools}
\usepackage{float}
\usepackage{siunitx}
\usepackage[english]{babel}
\usepackage{subfig}
\usepackage{color}
\usepackage{url}
\usepackage{hyphenat}
\usepackage{subfig}

\RequirePackage{flushend}
\RequirePackage{enumerate}
\RequirePackage[numbers,sort&compress]{natbib}
\RequirePackage{lineno}
\RequirePackage{fix-cm}
\RequirePackage{mathptmx}      
\RequirePackage{latexsym}
\RequirePackage[numbers,sort&compress]{natbib}
\RequirePackage[colorlinks,citecolor=blue,urlcolor=blue,linkcolor=blue]{hyperref}
\RequirePackage{mathrsfs} 
\RequirePackage[overload]{textcase} 
\RequirePackage{amssymb} 
\RequirePackage{upgreek}
\RequirePackage{dcolumn}
\RequirePackage{nth}

\newcommand{\Opera}{OPERA}
\newcommand{\numu}{\nu_{\mu}}

\newcommand{\antinu}{\overline{\nu}}

\newcommand{\nutau}{\nu_{\tau}}
\newcommand{\bmsigma}{\sigma \! \! \! \!\! \! \sigma}

\journalname{Submitted to EPJC}

\begin{document}

\title{First observation of a tau neutrino charged current interaction with charm production in the \Opera\ experiment}

\subtitle{OPERA Collaboration}

\author{
N.~Agafonova\thanksref{a1,a2}
\and
A.~Aleksandrov\thanksref{a3}
\and
A.~Anokhina\thanksref{a2}
\and
S.~Aoki\thanksref{a4}
\and
A.~Ariga\thanksref{a5}
\and
T.~Ariga\thanksref{a5,a6}
\and
A.~Bertolin\thanksref{a7}
\and
C.~Bozza\thanksref{a9}
\and
R.~Brugnera\thanksref{a7,a10}
\and
A.~Buonaura\thanksref{a3, a11}$\dagger$
% ...
}

\thankstext[$\star$]{c1}{Corresponding authors:\newline
marco.roda@liverpool.ac.uk,chiara.sirignano@unipd.it, gabriele.sirri@bo.infn.it}

\institute{INR - Institute for Nuclear Research of the Russian Academy of Sciences, RUS-117312 Moscow, Russia\label{a1}
\and
SINP MSU - Skobeltsyn Institute of Nuclear Physics, Lomonosov Moscow State University, RUS-119991 Moscow, Russia\label{a2}
\and
INFN Sezione di Napoli, 80125 Napoli, Italy\label{a3}
\and
Kobe University, J-657-8501 Kobe, Japan\label{a4}
\and
Albert Einstein Center for Fundamental Physics, Laboratory for High Energy Physics (LHEP), University of Bern, CH-3012 Bern, Switzerland\label{a5}
\and
Faculty of Arts and Science, Kyushu University, Japan\label{a6}
\and
INFN Sezione di Padova, I-35131 Padova, Italy\label{a7}
\and
Dipartimento di Matematica dell'Universit\`a di Salerno and ``Gruppo Collegato''  INFN, I-84084 Fisciano (Salerno), Italy\label{a9}
\and
Dipartimento di Fisica e Astronomia dell'Universit\`a di Padova, I-35131 Padova, Italy\label{a10}
\and
Dipartimento di Fisica dell'Universit\`a Federico II di Napoli, I-80125 Napoli, Italy\label{a11}
% ...
}


\date{}

\maketitle

\begin{abstract}

An event topology with two secondary vertices compatible with the decay of short-lived particles was found in the analysis of neutrino interactions in the \Opera\ target.
The observed topology is compatible with tau neutrino charged current (CC) interactions with charm production and neutrino neutral current (NC) interactions with $c\overline{c}$ pair production.
However, other processes can mimic this topology.
A dedicated analysis was implemented to identify the underlying process. A Monte Carlo simulation was developed and complementary procedures were introduced in the kinematic reconstruction. 
A multivariate analysis technique was used  to achieve an optimal separation of signal from background.
Most likely, this event is a $\nutau$ CC interaction with charm production, the tau and charm particle decaying  into 1 prong and 2 prongs, respectively. The significance of this observation is evaluated.

\end{abstract}

\newcommand{\tabtracksandvertices}{

\begin{table}[htb]
\centering
\addtolength{\tabcolsep}{-4pt}
\caption{Track slopes $p_{x}/p_{z}$ and $p_{y}/p_{z}$  at film 32 and their impact parameters (IPs), evaluated assuming a single vertex topology ($V_{5p}$) and a double vertex one. Errors on IP values are evaluated to be of the order of 0.6 $\mu \mathrm{m}$. 
The interaction is located well inside the brick.
The coordinates of vertices $V_I$ and $V_{II}$ are reported  in \tablename~\ref{tab:vertices}.
\label{tab:tracks_and_vertices}}

\begin{tabular}{ccc|c|cc}
 & & & \textbf{Single Vertex  IP}  ($\mu \mathrm{m}$) & \multicolumn{2}{c}{\textbf{Double vertex IP }} ($\mu \mathrm{m}$) \\
Track & $p_{x}/p_{z}$ & $p_{y}/p_{z}$ & w.r.t. $V_{5p}$ & w.r.t. $V_I$ &  w.r.t. $V_{II}$ \\
\hline
 1 & -0.230 & -0.275 &  8.3 & 36.2 &  0.2  \\ 
 2 &  0.121 & -0.144 &  8.8 &  1.0 &  6.5  \\
 3 &  0.349 & -0.036 &  4.8 & 25.9 &  0.2  \\ 
 4 & -0.003 &  0.088 & 13.0 &  1.5 & 20.4  \\
 5 & -0.003 & -0.025 &  5.1 &  2.2 &  9.6  \\

\end{tabular}
\end{table}
}

\newcommand{\tabvertices}{

\begin{table}[b]
\centering
\addtolength{\tabcolsep}{-4pt}
\caption[Vertex positions]{Position of the reconstructed vertices inside the brick evaluated with respect to the primary vertex $V_I$. A particle fully contained in the lead (non-visible) is associated with vertex $V_I$ and it is the parent of vertex $V_{II}$.}\label{tab:vertices}
\begin{tabular}{ccccccc}
Vertex & Type & Parent & Daughters & $x$ ($\mu$m) & $y$ ($\mu$m) & $z$ ($\mu$m) \\
\hline
$V_I$ & primary & - & 2, 4, 5, non-visible & 0 & 0 & 0 \\
$V_{II}$ & secondary  & non-visible & 1, 3  & 8 & -8 & 102 \\
$V_{III}$ & kink & 4 & 6 & -4 & 105 & 1155 
\end{tabular}
\end{table}
}

\newcommand{\tabphotons}{

\begin{table}[b]
\caption{Electromagnetic showers features. 
Reconstructed energies were estimated from the shower track multiplicity.
\label{tab:photons}}
\centering
\begin{tabular}{r|cc}
Shower ID & $\gamma_1$ & $\gamma_2$ \\
\hline
Starting film & 35 & 41 \\
$\theta_x$ (rad) & 0.050 & 0.011 \\
$\theta_y$ (rad) & 0.122 &  0.085 \\
IP$_{\mbox{I}}$ ($\mu$m) & $30 \pm 22$  & $40 \pm 23$ \\ 
IP$_{\mbox{II}}$ ($\mu$m) & $28 \pm 22$  & $40 \pm 23$ \\
IP$_{\mbox{III}}$ ($\mu$m) & $0.9 \pm 2.0$  & $40 \pm 11$ \\
Opening angle (rad) & 0.027 & 0.029 \\
Energy (GeV) & $7.1 \pm 1.7$  & $5.3 \pm 2.2$
\end{tabular}
\end{table}
}

\newcommand{\tabtracksmomentum}{

\begin{table}
\caption[Particles momenta]{Particle momenta reconstructed by the multiple Coulomb scattering method. \label{tab:tracks_momentum}}
\centering
\begin{tabular}{ccc}
Track ID & $p$ best fit (GeV/$c$) & 68\% C.L. $p$ range  (GeV/$c$) \\
\hline
1 & 2.1 & $\left[ \, 1.6 \, , \, 3.1 \, \right]$ \\
3 & 4.3 & $\left[ \, 3.1 \, , \, 7.1 \, \right]$ \\
5 & 0.54 & $\left[ \, 0.45 \, , \, 0.68 \, \right]$ \\
6 (daughter) & 2.7 & $\left[ \, 2.1 \, , \, 3.7 \, \right]$ 
\end{tabular}
\end{table}
}

\newcommand{\tabexpectedevents}{

\begin{table}[htbp]
\caption[Expected number of events with two secondary vertices]{Expected number of events with two secondary vertices as selected by the analysis. Parameters used to obtain the oscillated $\nu_\tau$ flux are: $\sin^2 2\theta_{13} = 9.3 \times 10^{-2}$, $\sin^2 2\theta_{23} = 1.0$, $\Delta m^2_{32} = 2.44 \times 10^{-3} \, \mathrm{eV}^2$, $\Delta m^2_{21} = 7.5 \times 10^{-5} \,  \mathrm{eV}^2$~\cite{pdg}. \label{tab:expected_events}}
\centering
\begin{tabular}{ll|c}
\multicolumn{2}{c|}{Samples} &  Expected number of  events ($10^{-3}$)  \\
 \hline
 & $\nutau$ CC + charm 
                            & $44.5$ \\
\hline 
 &  $\nu$ NC + $c\bar{c}$ pair & $12.6$ \\
 & $\numu$ CC + two 2ry       & $ 4.0$ \\
 & $\numu$ CC + charm + 2ry   & $20.5$ \\
 & $\nu$ NC + two 2ry         & $ 3.8$ \\
 & $\nutau$ CC +2ry           & $ 9.0$ \\\hline
 \multicolumn{2}{c|}{\textbf{Total} }               & $94.4$

\end{tabular}
\end{table}

}

\newcommand{\figoperadetector}{

\begin{figure}
\centering
\includegraphics[width=1\columnwidth]{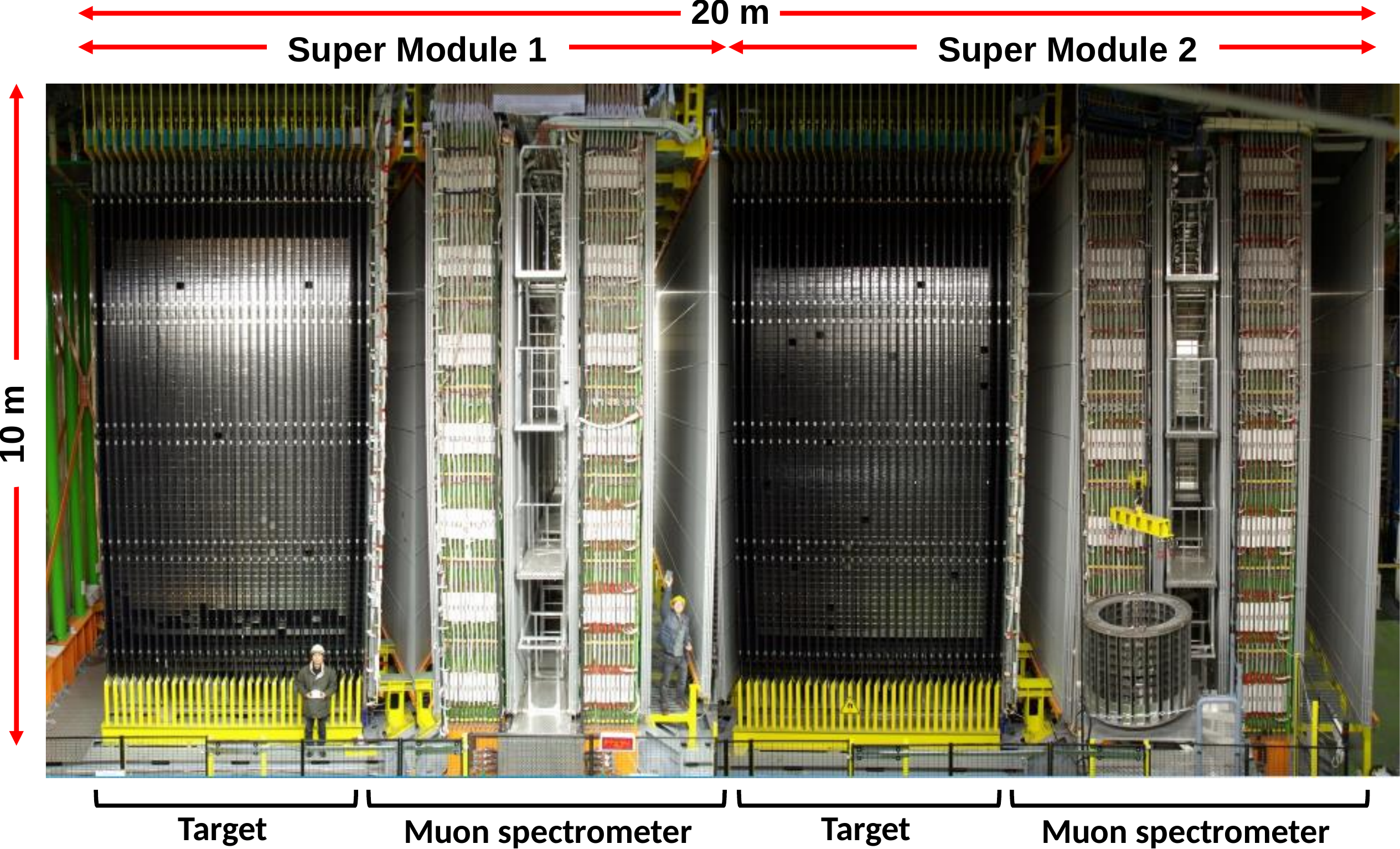}
\caption{Side view of the \Opera\ detector. The neutrino beam was coming from the left. 
The upper horizontal lines indicate the two identical supermodules (SM1 and SM2). 
\label{fig:opera_detector}}
\end{figure}

}

\newcommand{\figip}{

\begin{figure}
\centering
\includegraphics[width=.40\textwidth]{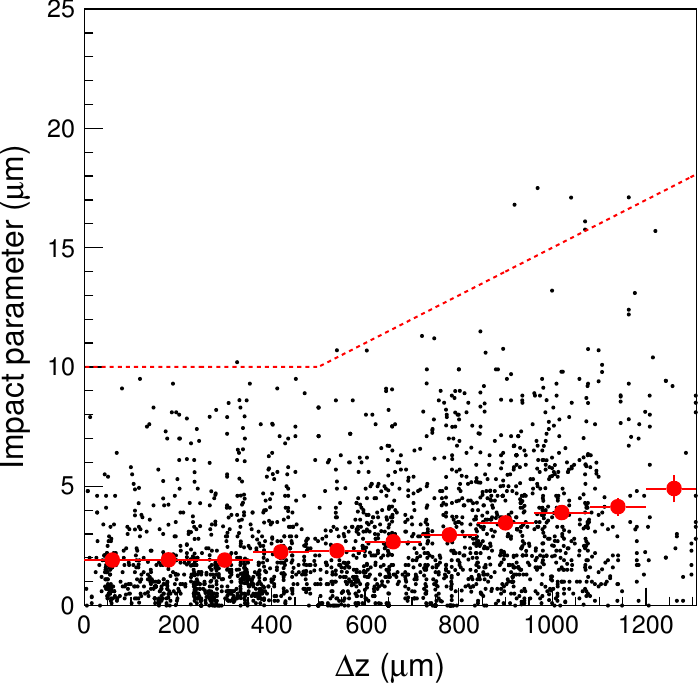}
\caption{Primary track impact parameters as a function of the longitudinal distance from the neutrino interaction vertex (2008--2009 data).
The red bullets show the average value for each bin. 
The dotted red line represents the lower limit applied to select decay candidates. \label{fig:ip}}
\end{figure}

}

\newcommand{\figeledet}
{

\begin{figure*}[t]
\centering
\includegraphics[width=1.6\columnwidth]{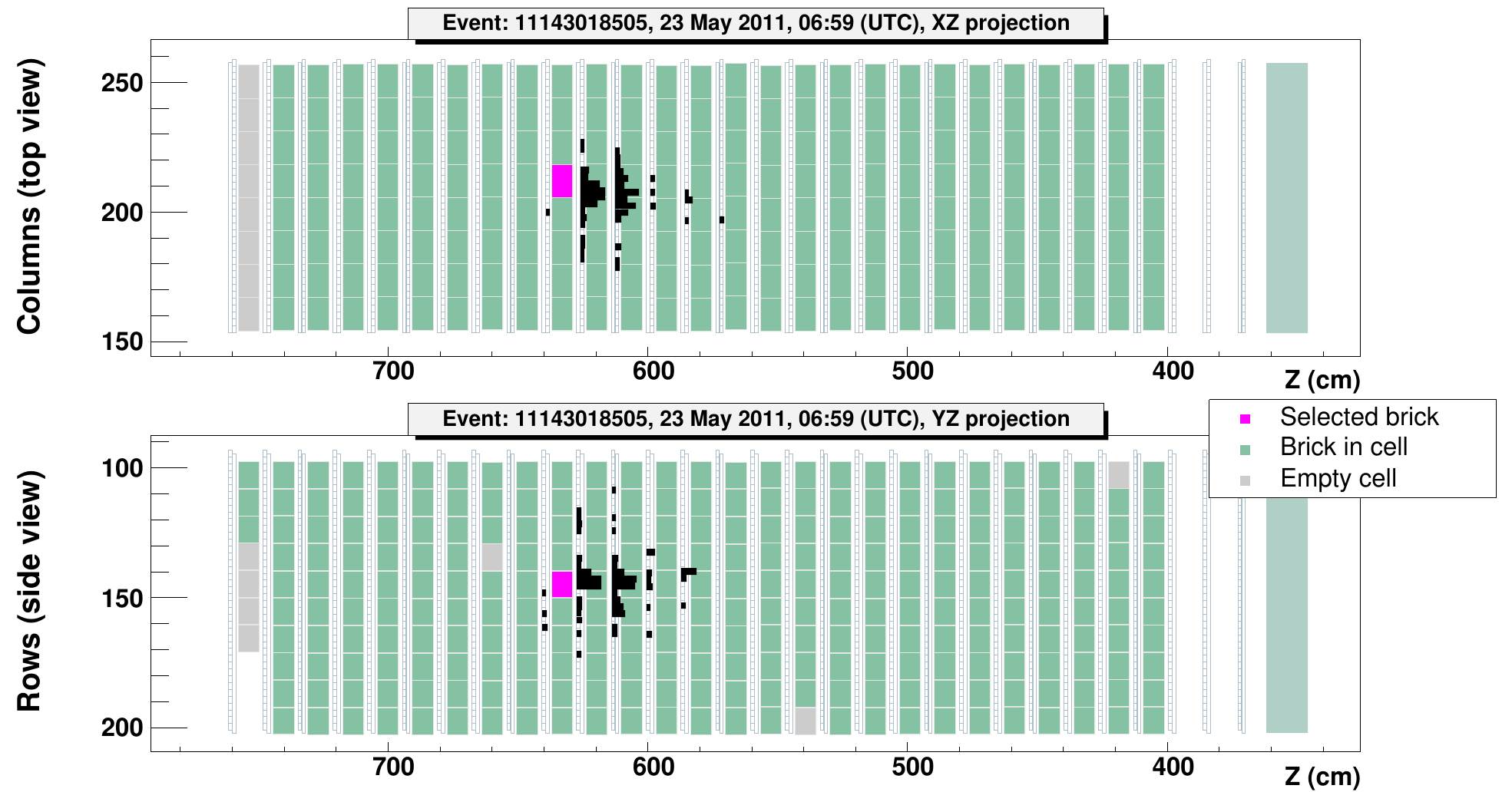}
\caption{Electronic detector display of the event. 
CNGS neutrinos travel from left to right. The angle between the
neutrino direction and the Z-axis projected into the XY plane is 58 mrad. Black bars indicate TT hits over threshold and are proportional to the released energy. Pink boxes show the most probable brick selected to search for the neutrino interaction vertex. The selected brick is located in wall 12 of the first SM.\label{fig:eledet}}
\end{figure*}
}

\newcommand{\figtomography}
{

\begin{figure*}[t]
\centering
\includegraphics[width=1.75\columnwidth]{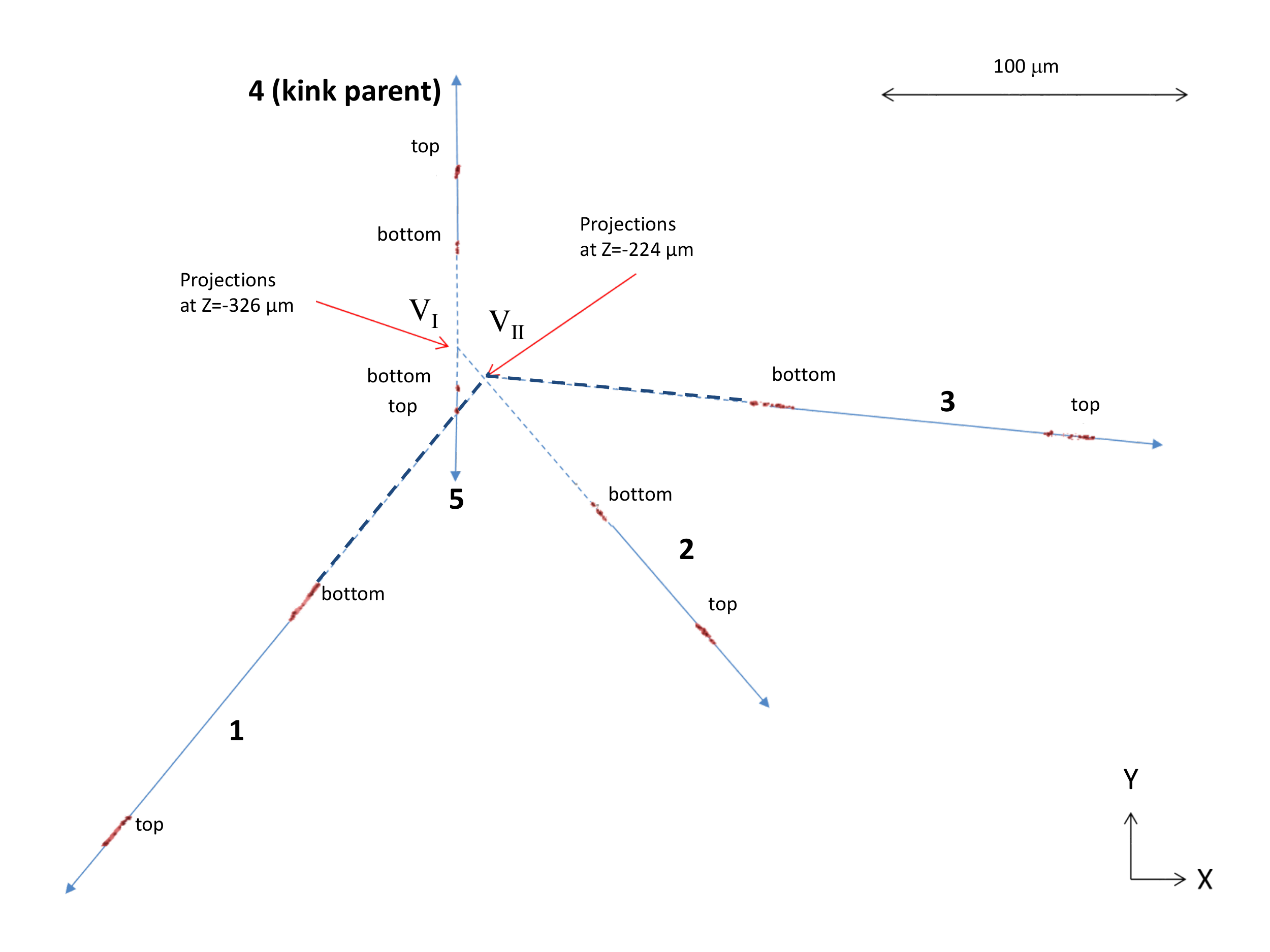}
\caption{Superposition of several tomographic emulsion images taken at different depths along film 32. Images are processed 
to show only grains related to the events.
Each track is composed by two sets of
grains, one for each sensitive layer (upstream, downstream) of film 32. X and Y are transverse coordinates while the Z axis is in the neutrino beam direction. 
Reconstructed tracks are shown with solid lines, the extrapolations are represented with segmented lines. According to the most probable topology of the event, tracks attached to Vertex $I$ are extrapolated with a dotted line while tracks attached to Vertex $II$ are extrapolated with a dashed line. Vertices $I$ and $II$ are highlighted with red arrows.
\label{fig:tomography}}
\end{figure*}

}

\newcommand{\figeventsideview}
{

\begin{figure}
\centering
\includegraphics[width=1.0\columnwidth]{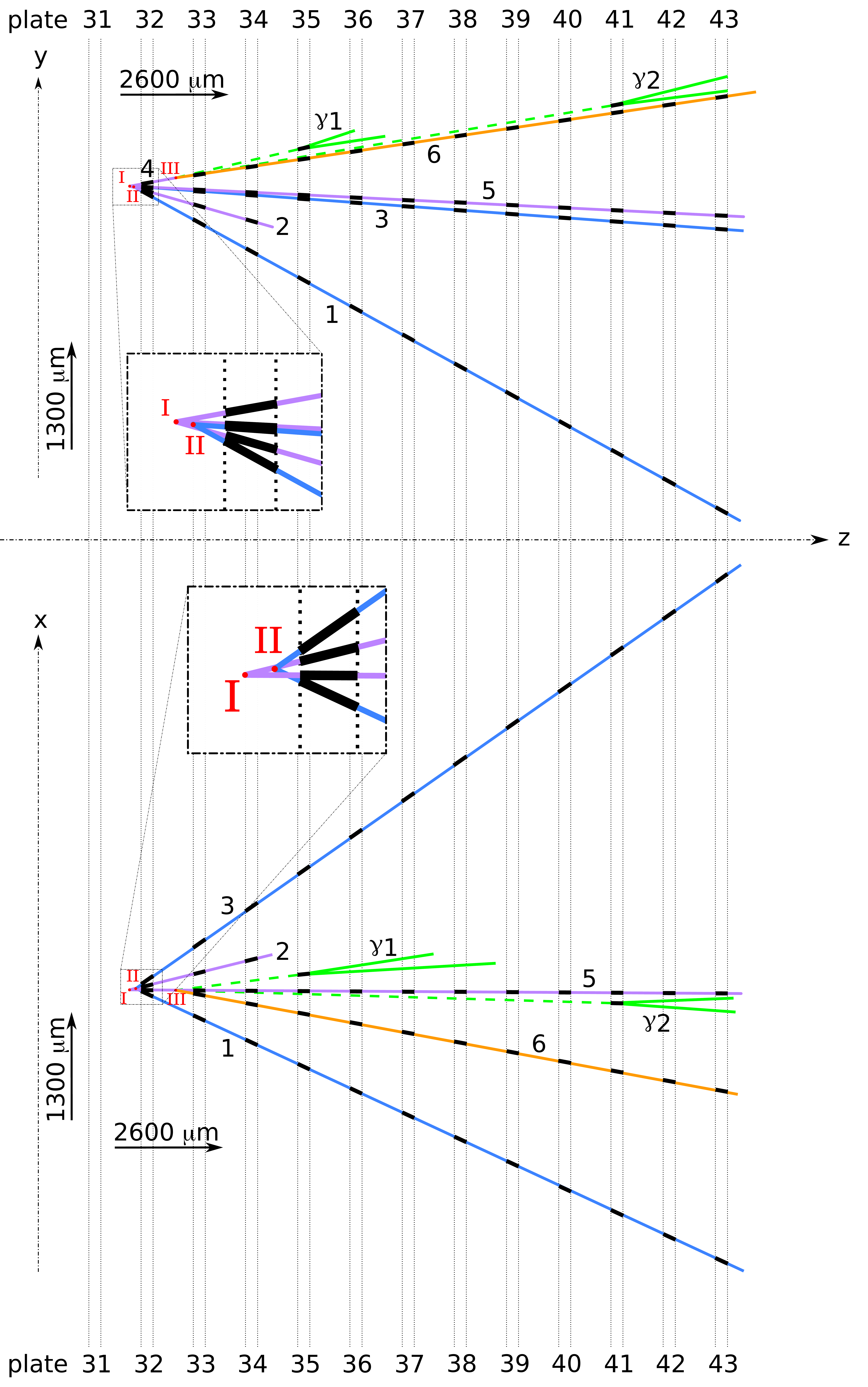}
\caption[Event top and side views around the primary vertex]{Projected views, in film 31 to 43, of the event in the $YZ$ (upper plot) and $XZ$ (bottom plot) planes.Tracks measured in single emulsion films are represented in black, while globally reconstructed tracks are represented using coloured lines: purple for tracks coming from the primary $V_I$, blue for tracks coming from $V_{II}$ and orange for the daughter of vertex $V_{III}$.
Photon directions are reported with green dashed lines until their first electron-positron pair. Continuous green lines represent cones containing the electromagnetic showers (not to scale in $X$,$Y$).

\label{fig:event_side_view} }
\end{figure}

}

\newcommand{\figshower}
{

\begin{figure}
\centering
\includegraphics[width=1\columnwidth]{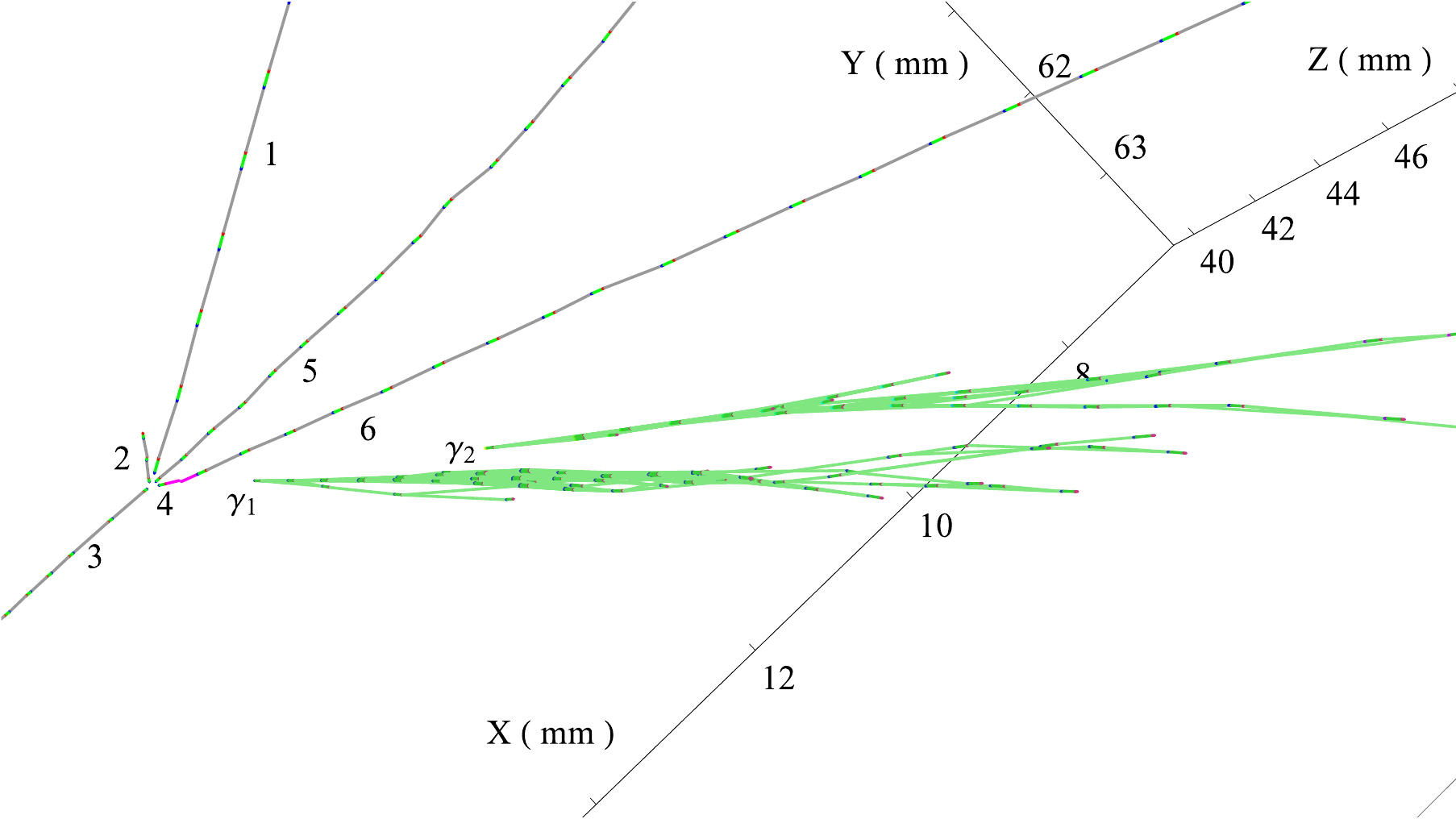} 
\caption{Reconstruction of the electromagnetic showers associated to the event.\label{fig:shower}}
\end{figure}

}

\newcommand{\figmlpspectrum}{

\begin{figure}[tbph]
\centering
\includegraphics[width=1\columnwidth]{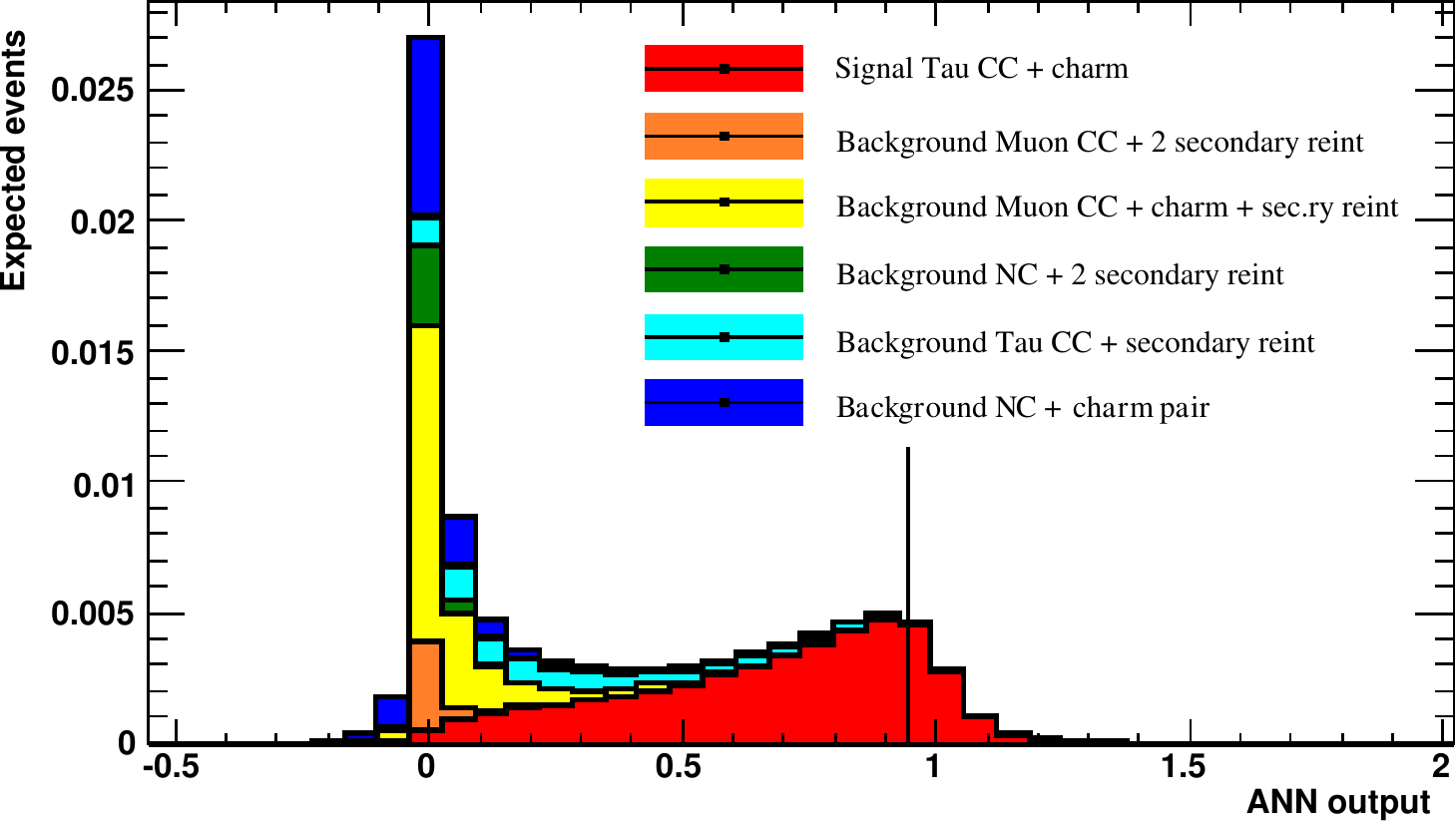}
\caption{Distribution of the ANN output variable. The weighted contribution of each process listed in \tablename~\ref{tab:expected_events} is shown with a different color.  
The vertical black line represents the ANN output for the event. \label{fig:mlp_spectrum}}
\end{figure}

}

\section{Introduction}

Charmed hadron production in neutrino interactions has been studied in two ways: dilepton searches in calorimeter detectors~\cite{charm2_dimuon} and  identification of charm decay topologies in nuclear emulsions~\cite{Burhop,charm1979,chorus_charm,Ushida,decay_search,double_charm_chorus,2charm_CHORUS}. Emulsion-based experiments allow a highly detailed reconstruction of the event topology, such that background can be reduced by a factor $10^{4}$~\cite{double_charm_chorus}. Background arises from pions and kaons decaying in flight or hadron interactions without any visible nuclear break-up. 

The \Opera\ experiment~\cite{opera_detector} was designed to observe $\numu \rightarrow \nutau$ oscillations in the CERN to Gran Sasso (CNGS) $\numu$ beam~\cite{CNGS} by the detection of tau leptons  produced in $\nutau$ CC interactions.
The experiment has been searching for neutrino interactions with one secondary short-lived particle as a signature of the $\tau$ lepton. OPERA reported in 2015 the discovery of $\nutau$ appearance in a muon neutrino beam~\cite{opera_discovery}, later extended to a significance of 6.1 $\sigma$ with its final data sample~\cite{opera_2018_PRL}. 

An interesting muon-less event with two secondary vertices was observed in the target of the \Opera\ detector. Both vertices can be interpreted as short-lived heavy particle decays.
Such an event can originate, at the CNGS energy, either from a $\nu_\tau$ CC interaction with charm production or from a $\nu$ NC interaction with $c\overline{c}$ production.
The first process is foreseen in the Standard Model but it was never been directly observed before, while the CHORUS experiment observed three events with $c\overline{c}$ production in $\nu$ NC interactions~\cite{double_charm_chorus}. The expected number of such events in OPERA is smaller than one.

In this paper the analysis and interpretation of this event is reported.
After a brief description of the apparatus (section \ref{sec::opera}), the event measurement and analysis are reported in sections \ref{sec::event_description} and \ref{sec::analysis}, respectively. The statistical significance of the observation is discussed in section \ref{sec::results}.

\section{The \Opera\ experiment\label{sec::opera}}

\figoperadetector
The detector was located at the LNGS underground laboratory and was exposed to the CNGS beam.
The experiment profited from a 730~km long baseline and the average neutrino energy was 17~GeV.
The beam exposure started in 2008 and ended in 2012, $1.8\times10^{20}$ protons on target were collected. $19\,505$ neutrino interactions were recorded in the fiducial volume of the detector target.

\subsection{The detector}
In order to observe and fully reconstruct decay topologies of short-lived particles,  a spatial resolution at the micrometer scale is required.
The target consisted of lead plates inter-spaced with nuclear emulsion films acting as high accuracy tracking devices, a configuration also known as Emulsion Cloud Chamber (ECC).\newline
The target was segmented into $150\,000$ units (bricks), each consisting of 57 nuclear emulsion films alternating with 56 ~1 mm thick lead plates.
Emulsion films were made of 2 emulsion layers, each 44 $\mu \mathrm{m}$ thick coated on both sides of a 205 $\mu \mathrm{m}$ transparent plastic base. 
The brick cross section was 103~mm~$\times$~128~mm; its thickness was 7.5~cm corresponding to about 10 radiation lengths. The brick mass was 8.3~kg. 
The achieved spatial resolution was $\sim 1 \, \mu \mathrm{m}$ and the angular resolution was $\sim 2$~mrad~\cite{opera_detector}.
Charged particle momentum is measured by Multiple Coulomb Scattering (MCS) in the lead plates~\cite{multiple_scattering}.
A changeable sheet (CS) doublet consisting of a pair of emulsion films is attached on the brick downstream face~\cite{CS_paper} as an interface between bricks and electronic detectors.

The active mass of the target  amounted to $1.25$ kton.
Bricks were housed in a modular detector structure made of two identical Super Modules 
(SM) \cite{opera_detector}. 
Each SM was composed of a target section and a muon spectrometer, as shown in \figurename~\ref{fig:opera_detector}.
In each SM, the bricks were arranged in 29 vertical walls orthogonal to the beam direction and alternated with electronic detectors~\cite{electronic_detectors} consisting of 
two orthogonal planes of plastic scintillators for each wall, called Target Tracker system (TT). 
TT planes were made up of scintillator strips 2.6~cm wide and 1~cm thick.
The TT was used to select the brick in which the neutrino interaction occurred. 
It also provided muon identification and an estimation of the energy deposited by hadronic and electromagnetic cascades.
The spectrometers were designed to measure the charge and the momentum of muons~\cite{opera_detector}.

\subsection{Event reconstruction\label{sec::event_reconstruction}}

The emulsion data taking is performed by fast automatic scanning systems, based on microscopes equipped with a computer-controlled motorised stage and a digital camera mounted on top of a dedicated optical 
system. 
Track recognition in emulsion films is performed on 16 tomographic images, grabbed at equally spaced depth levels through the 44 $\mu \mathrm{m}$ sensitive
layer~\cite{high_speed_tracking,scanning_hardware,japanese_scanning_system,european_scanning_system}.

The first step of the event reconstruction is the location of the primary neutrino interaction inside the brick~\cite{opera_detector}. 
The vertex location procedure in a brick starts from a set of predictions provided by the electronic detectors that are confirmed in the CS films.
Then, the tracks of secondary particles produced in the neutrino interaction are followed back in the brick, film by film, from the most downstream one to the interaction point
from where they originate. 
Whenever a track is not found in three consecutive films, a $1$~cm$^2$ surface  is scanned in each of the 5 films upstream and 10 films downstream of the last observed track segment in order to fully reconstruct the event.

In the decay search procedure~\cite{decay_search}, secondary vertices are searched for and all the selected tracks are double checked by manual measurements. The signature of a decay topology is the observation of a significant impact parameter (IP) with respect to the primary vertex.

The precision obtained in the vertex position is affected by particle scattering in lead plates that was evaluated by Monte Carlo (MC) simulations.
\figurename~\ref{fig:ip} shows the impact parameters of reconstructed primary tracks as a function of the primary vertex depth in lead ($\Delta$z).
If secondary vertices are found, a full kinematical analysis is performed combining the measurements in nuclear emulsions with data from the electronic detectors.

The appearance of the $\tau$ lepton is identified by the detection of its characteristic decay topologies, either in one prong (electron, muon or hadron) or three prongs.
Kinematical selection criteria are applied according to the decay channel \cite{opera_first_candidate,second_tau_paper}.

The detection and reconstruction efficiencies are evaluated by MC simulations~\cite{decay_search}.

\figip

\section{Description of the tau neutrino candidate event with charm production\label{sec::event_description}}

\figeledet

The event was recorded on May 23$^{\mbox{rd}}$, 2011 in the most upstream SM. 
The event display is shown in \figurename~\ref{fig:eledet}:
the number of fired TT planes is 9.
No muon track is reconstructed by the electronic detectors, thus the event is tagged as \textit{0-muon}. It is worth mentioning here that a track is tagged as a muon either 
if the product of its length and the density along its path is larger than 660 g/cm$^{2}$ or if the total number of electronic detector planes crossed is larger than 19. The energy reconstructed by the TT is equivalent to $20 \pm 6$~GeV~\cite{electronic_detectors}.

\subsection{Event topology\label{sec:topo}}

The neutrino interaction occurred in brick 77152 which was selected exploiting TT data using the procedure described in~\cite{electronic_detectors}. 
The analysis of the CS doublet reveals a converging pattern of 27 tracks. 
Out of them, 11 are found also in the most downstream film of the brick (plate 57). These tracks are clustered in a region of a few hundreds of micrometers, indicating an electromagnetic activity possibly related to the neutrino interaction.
All tracks are followed upstream in the brick: most of them are just few emulsion films long. 
By visual inspection these are confirmed as being part of an electromagnetic shower. 

In the location and decay search procedures (see section \ref{sec::event_reconstruction}), a primary convergence point of 5 tracks is found in film 32.

The reconstructed slopes are reported for each track in  \tablename~\ref{tab:tracks_and_vertices}.
The neutrino interaction point is located in the lead plate between emulsion films 31 and 32. 
As the impact parameter (IP) of track 4 w.r.t. the primary vertex is larger than the 10 {$\mu$}m threshold, a 5-prong primary vertex topology $(V_{5p})$ is discarded. 
The topology selected by the reconstruction algorithm \cite{decay_search} is a double vertex event with the primary neutrino vertex $(V_I)$ formed by tracks 2, 4 and 5, and a secondary vertex $(V_{II})$ formed by tracks 1 and 3 (see \figurename~\ref{fig:tomography}).

\tabtracksandvertices

Two additional measurements are performed, both yielding a better resolution than the standard one: i) manual measurement with a higher magnification objective and ii)~high-resolution automatic image acquisition and analysis.
In the first case, the tracks are measured in plate 32 and 33 under a 100 times magnification objective mounted on the microscope; thus achieving a $0.3$~$\mu$m resolution on the film transverse coordinates (X,Y) ~\cite{scanning_hardware}.
In the second case an improved scanning procedure based on emulsion images acquired with 1~$\mu$m pitch tomography, detailed in~\cite{showereco}, is applied. 
By this precise 3D tomographic technique it is possible to achieve a spatial resolution of 0.1~$\mu$m and an angular resolution of 1~mrad. 
Such accuracy allowed reducing track reconstruction from intrinsic emulsion background to less than 5~\%  while keeping track reconstruction efficiency close to 100~\%.

\figtomography

The primary vertex is reconstructed $(581.8 \pm 0.4) \, \mu$m upstream with respect to the downstream surface of film 32. 

A particle fully contained in the lead (non-visible), emitted at the primary vertex with angles  (0.086,~0.077)~rad and joining the secondary vertex $V_{II}$ has a flight length of $(103.2 \pm 0.4)~\mu$m. 

Track~4 (labelled as \textit{kink parent} in \figurename~\ref{fig:tomography}) exhibits a kink topology ($V_{III}$) between films 32 and 33. The minimum distance between track 4 and track 6 (\textit{daughter}) emerging from the kink, is $(0.9 \pm 0.4) \, \mu$m. 
The kink angle is $(95\,\pm\,2)$~mrad; the parent flight length is $(1174 \pm 5) \, \mu$m. A scheme of the full event is shown in \figurename~\ref{fig:event_side_view}. 

\tabvertices

All tracks reconstructed at films 32 and 33 are followed down in the brick in order to estimate their momenta.

\figeventsideview

Track~2 stops at film 34;
track 3 undergoes a re-interaction at film 53, while tracks~1,~5 and~6 reach the CS films. The coordinates of the three vertices are listed in  \tablename~\ref{tab:vertices}.

Two $e^+e^-$ pairs
are identified in films 35 ($\gamma_1$) and 41 ($\gamma_2$). An image based procedure is applied to identify and reconstruct the electromagnetic showers~\cite{showereco}. An additional image data taking with 1 micron $Z$ pitch is performed in a cone of 400~mrad aperture around the slope of the primary photon ($e^+ e^-$ mean slope), starting from film 31 down to film 57 in the brick. 
All tracks in the volume are reconstructed using the 3D clustering algorithm.
The main features of the reconstructed showers are listed in \tablename~\ref{tab:photons} and shown in \figurename~\ref{fig:shower}.

\tabphotons

\figshower

The most downstream shower tracks are reconstructed in the CS films.
Given the accuracy of the reconstruction, $\gamma_1$ is most probably attached to $V_{III}$, while $\gamma_2$ may emerge from any vertex ($V_I$, $V_{II}$ or $V_{III}$), see IPs in \tablename~\ref{tab:photons}. 
In the next step of the analysis the photons origin vertex is not taken into account since the classifier procedure relies only on the total visible electromagnetic energy.
In order to reduce hadron interaction background, a dedicated scanning system with high efficiency at large angles is also used~\cite{hadrware_performance} to search for nuclear fragments around each vertex and around the stopping point of Track 2, within a $| \tan \theta | < 3$ acceptance window. No nuclear fragments are detected.
Independent analyses were performed in three different scanning laboratories confirming all the results.

\subsection{Event kinematics \label{sec:kinematic}}

Momenta of tracks 1, 3, 5, 6 are estimated using the MCS method. 
The alignment uncertainties are evaluated from angular and position residuals of a sample of tracks penetrating the  entire scanned volume.
All measurements are performed using high-resolution images taken with the acquisition system described in section ~\ref{sec:topo}. The angular and position resolutions are 2.4 mrad and 0.6 $\mu$m, respectively.
Results are shown in \tablename~\ref{tab:tracks_momentum}. 

\tabtracksmomentum

Track 2 is observed only in three emulsion films and it is identified as a Minimum Ionizing Particle (MIP) by counting the grains along the track.
The momentum estimation by range~\cite{nist_range} discards the proton hypothesis. The particle is most likely a pion with an energy of 0.2 GeV.
An additional estimation is performed considering absorption processes. 
Pion absorption cross-section has a resonance at a kinetic energy of about 0.2~GeV in any material~\cite{pion_general_absorption_peak,pion_absorption_dependency}.
In this region, especially for heavy nuclei, this cross section 
is up to $\sim$~40~\% of the total cross section.
Under these assumptions, the momentum estimation for track~2 is $(0.31 \pm 0.08)$~GeV$/c$. This is the initial momentum of a pion which is absorbed after crossing 3~mm of lead and that has 
a kinetic energy of about $0.2$~GeV when absorbed. 
The uncertainty is evaluated assuming a uniform kinetic energy distribution: 
the minimum is the kinetic energy such that $\beta > 0.7$; while the maximum is 0.3~GeV, which is the endpoint of the absorption peak. $\beta=0.7$ is the minimum for a MIP according to the emulsion grain density~\cite{test_beam}. 

The energies of the electromagnetic showers, $\gamma_1$ and $\gamma_2$, are estimated by counting the tracks belonging to each shower. The procedure is calibrated with MC simulations, taking also into account background tracks~\cite{showereco}. The result is  $E_{\gamma_1} = ( 7.2 \pm 1.7 )$~GeV and $E_{\gamma_2} = ( 5.3 \pm 2.2 )$~GeV. 
Considering the absence of large angle scattering on track~6, $\gamma_1$ and $\gamma_2$ are not bremsstrahlung photons. Therefore tracks 4 and 6 can be neither positrons or nor electrons.

In conclusion, the event is identified as a neutrino interaction with two secondary vertices: $V_{II}$ has a 2-prong topology while $V_{III}$ is a kink originated by a primary charged particle.
The minimum invariant masses \cite{minimum_invariant_mass} at the secondary vertices were estimated to be $2.5 \pm 0.8$~GeV and $1.8 \pm 0.4$~GeV respectively.

\section{Event analysis\label{sec::analysis}}

The event described in this paper was one of the 10 events selected as $\nutau$ candidates in the     
OPERA final analysis on $\numu \rightarrow \nutau$ oscillation~\cite{opera_2018_PRL}. 
Nevertheless, given the presence of two short-lived particle decay candidates (the kink topology and the additional vertex) a further dedicated analysis was performed for the event classification. 

Two short-lived particle decays can be produced by the following processes:
\begin{itemize}
	\item $\nu_\tau$ CC interaction with charm production;
    \item $\nu$ NC interaction with $c  \overline{c}$ pair production.
\end{itemize}
Other processes mimicking this topology are:

\begin{itemize}
	\item $\nu_\mu$ CC interaction with a mis-identified muon and two secondary interactions.
	\item $\nu_\mu$ CC interaction with single charm production, a mis-identified muon and one secondary interaction;    
    \item $\nu$ NC interaction with two secondary interactions;
	\item $\nu_\tau$ CC interaction with one secondary interaction.
\end{itemize}

As secondary interaction, is meant either i) a hadronic interaction of a final state particle, ii) a decay of pions or kaons, or iii) a large-angle Coulomb scattering of hadrons or mis-identified muons.
Additional processes leading to similar topologies but with cross-sections smaller by at least one order of magnitude with respect to those considered in this analysis have been ignored.

The analysis is intended to establish the likelihood of the event with respect to different physics hypotheses. It is based on the distributions of kinematical variables obtained through a dedicated MC production.
Neutrino interactions are generated using GENIE~\cite{genie_MC}, except for charm pair production, simulated using HERWIG~\cite{herwig_MC}. 
The $\nu_\tau$ CC interaction cross sections used are in agreement with~\cite{nutau_charm_cs}.
Due to the high multiplicity of MIP tracks associated to the primary vertex for the event, only DIS interactions are taken into account. In total, about 300~million events are generated.

Particles from neutrino interactions are propagated in a 17 cm$^3$ volume of the brick using the Geant4 framework~\cite{geant4_1, geant4_2}, generating the primary vertex always at the same depth in lead as the one estimated for the event. 
The MCS is taken into account using a parameterisation based on the standard \Opera\ MC.
The hadron interaction simulations are validated using dedicated test beam data~\cite{test_beam}. 

For each process, the number of expected events is normalised to the $12\,352$  observed $\numu$ CC events with a primary vertex in the target section of the detector.
This strategy is applied in order to maximise the sample under analysis. The shape of the CNGS neutrino flux \cite{CNGS}, the oscillation probabilities and the cross sections are considered. 
The vertex location efficiency is determined according to a data-driven parameterisation.
The efficiencies related to the electronic detectors (brick selection, muon identification, muon momentum estimation) are evaluated using the standard MC with a parameterisation based on hadronic energy and muon momentum. Simulated events are selected regardless of the multiplicity at the primary vertex by requiring:
\begin{itemize}
\item no muon identified by the electronic detectors;
\item a one prong-like secondary vertex (1pr-like) with charged parent;
\item a two prong-like secondary vertex (2pr-like);
\item no nuclear fragments at any vertex.
\item daughter particles should not be electrons nor positrons.
\end{itemize}
Requiring a charged parent for the 1pr-like secondary vertex implies that the parent track has to be measured in at least one emulsion film. No kinematic cuts are applied. The total number of expected events matching this topology 
is $\sim0.1$ as shown in \tablename~\ref{tab:expected_events}.
\tabexpectedevents

\figmlpspectrum

A multivariate analysis is applied to the selected events and the signal to background discrimination is based on 12 kinematic variables as follows.

\begin{itemize}
\item for the entire event: i) \textit{total EM energy}, i.e. the sum of any reconstructed photon energy, regardless of the photon origin vertex;
ii) the \textit{angle} $\varphi$ between the parents of the 1pr-like and 2pr-like vertices in the transverse plane;
iii) the \textit{projection} in the transverse plane of the momentum at the primary vertex;
iv) the \textit{hadronic momentum}, i.e. the sum 
      of the primary track momenta excluding the
      two parents;

\item for the 1pr-like vertex: v) the \textit{daughter momentum}; vi) the \textit{daughter transverse momentum} with respect to the parent direction; vii) the \textit{flight length}; viii) the \textit{kink} angle between parent and daughter;

\item for the 2pr-like vertex: ix) the total \textit{daughters' momentum}; x) the total \textit{daughters' transverse momentum} with respect to the parent direction; xi) the \textit{flight length}; xii) the \textit{invariant mass} of the charged daughters.

\end{itemize}

In order to find the best method for the discrimination of the $\nu_\tau$ CC interaction with charm production, several algorithms have been tested: an Artificial Neural Networks (ANN) method \cite{tmva}, two kinds of Boost Decision Trees \cite{ada_boost} and the Fisher Discriminant \cite{fisher}. 
The best one turns out to be the ANN, whose output variable distribution is shown in \figurename~\ref{fig:mlp_spectrum}. According to this multivariate analysis the event can be classified as $\nutau$ CC interaction with charm production with rather high probability.

\section{Results\label{sec::results}}

The significance of the $\nutau$ CC interaction with charm production (\textit{signal}) observation is based on a frequentist hypothesis test using a likelihood ratio as a test statistic~\cite{profile_likelihood_ratio, likelihood_test}.

The ANN response $x$ plays the role of observable and its distribution, normalised to the expected number of events, is shown in \figurename~\ref{fig:mlp_spectrum}.
The shape of the distribution of $x$ is obtained by the sum of the contributions of each process listed in \tablename~\ref{tab:expected_events} and is described by:
\begin{equation}
   \sum_{i \in B} n_i  b_i(x) + \mu n_s s(x)   \label{eqn:no_syst_pdf}
\end{equation}
where index $i$ ranges over the background processes; $b_i(x)$ and $s(x)$ are the probability densities for background and signal, respectively; $n_i$ and $n_s$ are the expected number of events. The parameter $\mu$ (signal strength) is introduced as a scale factor on the number of signal events: $\mu=0$ corresponds to the background-only hypothesis and $\mu=1$ corresponds to the $\nutau$ CC interaction with charm production as predicted by \tablename~\ref{tab:expected_events}. 

The effect of uncertainties on the expected number of events are introduced as scale factors $f_i$ for each background process and $f_s$ for the signal.
They depend on 5 nuisance parameters: i) a 20~\% normalisation factor, dominated by the CNGS flux uncertainty~\cite{CNGS};
ii) a $20~\%$ uncertainty on the cross section of $\nu$ NC interactions with charm pair production;
iii) a $20\%$ uncertainty on the cross section of $\nutau$ CC interactions with single charm;
iv) a $6~\%$ uncertainty on $\nutau$ CC interactions cross section (without charm),  for $\nutau$ energies in the range of few tens of GeV \cite{nutau_cs};
v) a 30~\% uncertainty on the hadronic re-interaction rate, based on test beam results~\cite{test_beam}.
These nuisance parameters $\sigma_k$ are constrained by some uniform or Gaussian probability densities $g_k$.

Including systematic these terms and considering that the probability to observe N events  follows a Poisson distribution, the likelihood function is:
\begin{align}
 \mathcal{L} \left( \mu, \bmsigma  \right) 
 = 
 \mathrm{Pois} \left( N | \nu( \mu, \bmsigma ) \right) \prod_{j=1} ^{N} \Big[ \sum_{i \in B} f_i( \bmsigma ) \, n_i b_i(x_j) \nonumber\\
 + \mu f_s ( \bmsigma ) \, n_s s(x_j) \Big] \prod_{k=1} ^{5} g_k(\sigma_k)
 \;  \label{eqn:complete_pdf}
\end{align}

\noindent where $\bmsigma$ denotes the whole set of nuisance parameters, and $\nu( \mu, \bmsigma ) $ is the number of expected events.

In order to test which values of the signal strength $\mu$ are consistent with data, the profile likelihood ratio
$\lambda(\mu) = \mathcal{L}(\mu,\hat{\hat{\bmsigma}})/\mathcal{L}(\hat{\mu}, \hat{\bmsigma})$ is used~\cite{pdg}, where $\mathcal{L}(\hat{\mu}, \hat{\bmsigma})$ is the value of the likelihood at its maximum and $\hat{\hat{\bmsigma}}$ indicates the profiled values of the nuisance parameters $\bmsigma$, maximizing $\mathcal{L}$ for the given $\mu$. The distribution for the profile likelihood ratio $\lambda(\mu)$ is obtained with a sample of Monte Carlo pseudo-experiments generated according to the background-only hypothesis using RooFit RooStats libraries~\cite{roofit} and the ROOT framework~\cite{ROOT}.
For each pseudo-experiment, nuisance parameters $\sigma_k$ are randomly generated according to their PDFs $g_k$.

The probability that the background-only would produce events less likely compatible with the observed one is $(1.3\pm0.3)\times10^{-5}$.

This result refers to the observed topology, i.e. to the search for a single prong tau decay and a neutral charmed hadron decaying into 2 prongs. It provides evidence for the first observation of a  $\nu_\tau$ CC interaction with a charmed hadron.
The significance of this observation is {$4.0 \, \sigma$}.

Under the condition that the number of events is regarded as fixed, i.e. the Poissonian term is excluded from the likelihood, the p-value is $(6.6\pm0.1)\times10^{-3}$, corresponding to $2.5 \, \sigma$.
This estimate only relies on the event features
and it is independent on fluctuations of the total number of observed events.

The most likely interpretation is that $V_{II}$ is a charmed particle decay and $V_{III}$ is a tau lepton decay into a hadron.

\section{Conclusions}

A neutrino interaction was observed in the target of the OPERA detector having a rare topology: two secondary vertices within about 1~mm from the primary one were reconstructed. 
High-accuracy scanning procedures were applied and a dedicated analysis was set up.
Monte Carlo simulations were developed and additional procedures were introduced in the kinematic reconstruction. 
Multivariate analysis techniques were used to achieve an optimal separation between signal and background.

The event reported in this paper is the first observation of a $\nutau$ CC interaction with charm production candidate, tau and charm particles decaying into 1 prong and 2 prongs, respectively. The significance of this observation is {$4.0 \, \sigma$}. 

Under the condition that the number of events is regarded as fixed, i.e. the Poissonian term is not included in the likelihood, the significance of the observation is $2.5 \, \sigma$.

The observed process is foreseen in the Standard Model but it was never observed before. Our observation, although statistically limited to a single event, is compatible with the Standard Model expectation. Future experiments at CERN using neutrinos from proton collisions on a fixed target may study this
process with larger statistics~\cite{SHiP,LHCneutrini}.


\section*{Acknowledgements}
We acknowledge CERN for the successful operation of the CNGS facility and INFN for the continuous support given to the experiment through its LNGS laboratory.\newline We acknowledge funding from our national agencies: Fonds de la Recherche Scientifique-FNRS and Institut Inter Universitaire des Sciences Nucleaires for Belgium; MoSES for Croatia; CNRS and IN2P3 for France; BMBF for Germany; INFN for Italy; JSPS, MEXT, the QFPU-Global COE program of Nagoya University, and Promotion and Mutual Aid Corporation for Private Schools of Japan for Japan; SNF, the University of Bern and ETH Zurich for Switzerland; the Russian Foundation for Basic Research (Grant No. 12-02-12142 ofim), the Programs of the Presidium of the Russian Academy of Sciences (Neutrino Physics and Experimental and Theoretical Researches of Fundamental Interactions), and the Ministry of Education and Science of the Russian Federation for Russia, the National Research Foundation of Korea (NRF) Grant (No. 2018R1A2B2007757) for Korea; and TUBITAK, the Scientific and Technological Research Council of Turkey for Turkey. 
We thank the IN2P3 Computing Centre (CC-IN2P3) for providing computing resources.

\bibliography{bibliography}{}
\bibliographystyle{spphys}       

\end{document}